\documentclass[12pt]{iopart}
\usepackage{amssymb}
\usepackage[latin1]{inputenc}
\usepackage{amsfonts}
\usepackage{graphicx}
\usepackage{cite}
\usepackage{color}
\usepackage{bm}
\usepackage{dsfont}
\usepackage{hyperref}

\usepackage{ulem}  
\newcommand{\bra}[1]{\ensuremath{\left\langle#1\right|}}
\newcommand{\ket}[1]{\ensuremath{\left|#1\right\rangle}}
\newcommand{\dyad}[2]{\ensuremath{\left|#1\right\rangle\left\langle#2\right|}}
\newcommand{\bdyad}[3]{\left[\dyad{#1}{#2}\right]_{#3}}

\begin{document}

\title[Gaussian processes for parameter selection for Rydberg aggregates]{Gaussian processes for choosing laser parameters for driven, dissipative Rydberg aggregates}

\author{C D B Bentley and A Eisfeld}

\address{Max Planck Institute for the Physics of Complex Systems, N\"othnitzer Strasse 38, 01187 Dresden, Germany}
\ead{\mailto{cbentley@pks.mpg.de}, \mailto{eisfeld@pks.mpg.de}}

\begin{abstract}
To facilitate quantum simulation of open quantum systems at finite temperatures, an important ingredient is to achieve thermalization on a given time-scale.
We consider a Rydberg aggregate (an arrangement of Rydberg atoms that interact via long-range interactions) embedded in a laser-driven atomic environment.
For the smallest aggregate (two atoms), suitable laser parameters can be found by brute force scanning of the four tunable laser parameters.
For more atoms, however, such parameter scans are too computationally costly.
Here we apply Gaussian processes to predict the thermalization performance as a function of the laser parameters for two-atom and four-atom aggregates.
These predictions perform remarkably well using just 1000 simulations, demonstrating the utility of Gaussian processes in an atomic physics setting.
Using this approach, we find and present effective laser parameters for generating thermalization, the robustness of these parameters to variation, as well as different thermalization dynamics.
\end{abstract}

\section{Introduction}

One often encounters a situation where the outcome of an experiment depends on many control parameters that can be varied over a large range. Usually, one is interested in achieving a particular outcome.
This scenario emerges in both experimental and theoretical settings.

Achieving the target outcome can be a demanding task, in particular when the space of tunable parameters is high dimensional, and when there is no simple dependence of the outcome on the parameters.
If one is only interested in the optimal outcome, then various approaches have been developed for this task (e.g. the methods in~\cite{bv04,Stor97JGO,Duec90JCP}).
However, one is typically also interested in the 'stability' of the outcome, not (just) the parameter values that give the optimal outcome.
Additionally, one would like to know the regions of parameter space where one is close to the desired outcome.
We quantify 'closeness' to the target outcome by a \emph{cost function} (or simply \emph{cost}); when we speak about 'close to the desired outcome' we mean loosely that the result is below a certain value of the cost.
Similarly, when we speak about 'optimal' we mean the set of parameters that gives a result that is closest to the desired outcome; yielding the minimal cost.
For convenience, we will denote the manifold of the cost function over parameter space as the \emph{cost landscape}.

In the ideal case one would like to know the full cost landscape: the respective values of the cost for all choices of the tunable parameters.
However, analytic descriptions of the cost landscape can rarely be found for complex systems.
Also, brute force scans of the parameter space are often precluded by the time required to perform an experiment/calculation for a single set of parameters. 
One powerful approach is based on Gaussian processes (GPs)~\cite{RW}. 
GPs are not only able to search for an optimum but also provide information on the full cost landscape. 
This is performed by regression: a Gaussian process provides a prediction of the cost landscape by fitting known data, using Bayesian updates to prior assumptions for the model. 
This method can provide predictions of the full cost landscape from few data points, and has been applied to predict interatomic potentials~\cite{Bart10PRL,Bart15IJQC}, and the related kernel ridge regression method has also been applied to predict properties of atoms in molecules~\cite{Rupp12PRL,Rupp15JPCL}.

Knowledge of the predicted cost landscape also provides valuable information for choosing subsequent parameters for the experiment/simulation 'well', to use resources efficiently.
This is related to reinforcement learning, where previous 'policies of action' inform the following action policies, with some trade-off between exploring new policies and exploiting those that have worked well previously.
Reinforcement learning has been applied recently for quantum control~\cite{Buko17arX,Augu18arX,Fose18arX,Alba18arX} and even to design quantum optics experiments~\cite{Meln17PNAS}.
The idea of using known points in the cost landscape to choose subsequent parameters for evaluation is known as active learning~\cite{Settles2009}.
Active learning has been applied in physical settings to produce Bose-Einstein condensates~\cite{Wigl16SR25890}, and for materials design~\cite{Ju17PRX,Li17SR}.
Active learning is particularly useful when each simulation is computationally expensive.

\begin{figure}[tb]
	\centering
	\includegraphics[width=8cm]{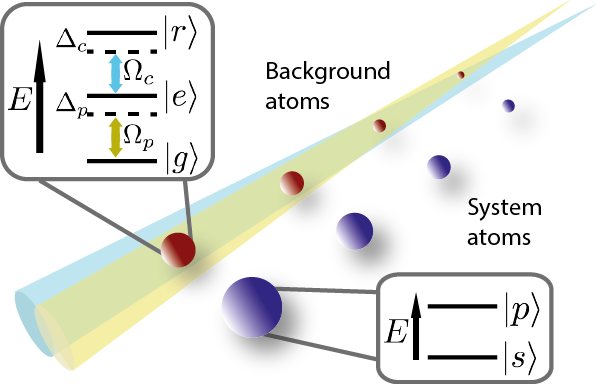}
	\caption{Setup: linear arrangements of system and background atoms, with lasers addressing the background atoms.  The atomic energy levels of system and background atoms are shown.
	The lasers drive the atomic transitions in the background atoms, and the Rabi frequencies $\Omega_p$, $\Omega_c$ and detunings $\Delta_p$, $\Delta_c$ are the control parameters for our setup.
	}
	\label{fig:setup}
\end{figure}

Here we consider such a problem in the context of using interacting Rydberg molecules for simulating open quantum system dynamics.
Rydberg systems have strong interactions, and can be optically addressed and positioned~\cite{Brow16JPB,Nogr14PRX,Schl11QIP}.
These properties can be exploited to simulate quantum systems~\cite{Wust18JPB,Weim10NP,Labu16N,Barr15PRL,Hagu12NJP,Lesa12PRL,Mulk07PRL}.
In addition, Rydberg atoms exhibit state-changing interactions similar to molecular interactions~\cite{Robi04PRA,Ditz08PRL}.
Since the parameters of Rydberg atomic systems are considerably simpler to control than the molecular counterparts, these systems are a promising setting for simulating molecular dynamics such as excitation transport in light-harvesting complexes.
Features necessary for simulating molecular systems have been demonstrated using Rydberg atoms, including tunable excitation transport~\cite{Gunt13S,Schon15PRL123005,Sche15PRL}, non-Markovian behaviour induced in the system~\cite{Genk16JPB}, as well as controllable thermalization~\cite{Schon18NJP013011}.
We will use the same setup as in~\cite{Schon18NJP013011} (related to the setups in~\cite{Schon15PRL123005,Genk16JPB}), as shown in figure~\ref{fig:setup}, with a laser-driven atomic environment.
Our setup has thus been used to demonstrate thermalization and flexible system dynamics, with remarkable control provided by the laser parameters.
In this work, we denote the tunable laser parameters (Rabi frequencies and detunings) our \emph{laser control parameters}, and \emph{parameter space} is the space of possible laser control parameters.

There are key remaining questions about thermalization and thermalization dynamics to show that quantum simulation of molecular systems is possible.
The first questions are about locating effective laser control parameters: which sets of parameters give rise to thermalization for a given setup?  How robust are these parameters to variation?
It is also important to understand the relationship between parameters and thermalization temperature: for parameters that generate thermal states, is there a smooth relationship between the parameters and thermal-state temperature?
Additionally, quantum simulation requires a system with sufficient size to mimic the target system.
This typically requires more than a couple of system atoms, so we also have questions about scalability: do the parameters that produce thermalization change as we scale the number of atoms in the system? If so, how?
Finally, approaching a thermal state in the 'right' way - the same way as a target molecular system - is just as important as the thermalization itself.
This leads to important questions about the thermalization dynamics: are there different sets of parameters that generate thermalization, with different thermalization dynamics? By varying parameters, what control do we have over the timescale of thermalization relative to interaction timescales within the system?

Given the significance of these open questions in our setup, our outcome of interest in this work is the thermalization of the Rydberg system atoms.
Before we can approach the open questions, however, we require a method for exploring the cost landscape.
Here the cost is a measure of the distance between the actual state and the target thermal state, and the cost landscape is over the laser control parameters.

In this paper, we will focus on the primary problem of exploring the cost landscape for a multidimensional parameter space, and costly simulations of our setup.
This investigation will provide insights into some of the various questions we have posed to facilitate quantum simulation of open systems, and provide a methodology for approaching any of these questions.

The structure of the manuscript is as follows: first we introduce our setup, methodology and performance measures in section~\ref{sec:setup}.
We then present results from scanning the parameter space for a dimer (two atom) aggregate system in section~\ref{sec:scan}.
GPs are applied to predict the cost landscape for a dimer system in section~\ref{sec:dim}.  Here the predicted landscape is presented and analysed using our performance measures.
Then, in section~\ref{sec:quad}, the cost landscape is predicted using GPs for a quadromer system, and we discuss the physical utility and the performance of the prediction.
Finally, we present our conclusions in section~\ref{sec:conc}.

\section{Setup, basic definitions and methods} \label{sec:setup}

\subsection{The system and its tunable parameters} \label{sec:systun}

We consider the setup shown in figure~\ref{fig:setup}.
This model has been discussed in detail in Refs~\cite{Schon15PRL123005,Schon18NJP013011}.

The system is a Rydberg $k$-mer of $k$ atoms interacting via resonant-dipole-dipole interactions.  
These atoms remain in one of two Rydberg states: the lower-energy $\ket{s}$ state and the excited $\ket{p}$ state.
The Hamiltonian for the system is given by:
\begin{equation}\label{eq:Hfullsys}
 \mathcal{H}_\mathrm{sys} = \sum_{n\neq m} W_{nm}\dyad{\pi_n}{\pi_m},
\end{equation}
where $\ket{\pi_n} = \ket{s \cdots s p s \cdots s}$ represents the system when a single $\ket{p}$ excitation is localized at atom $n$.  
 A single excitation is shared between the system atoms in our setup. 
The resonant dipole-dipole interaction $W_{nm}=C_3/(\bm{R}_n-\bm{R}_m)^3$, where $\bm{R}_n$ is the position of atom $n$ and $C_3$ is a state-dependent coefficient.
The system atoms are arranged linearly, with separation distance $d$ between atoms.

The environment is composed of laser-driven atoms.  These atoms are a distance $\delta$ from the system atoms, such that the vectors along $d$ (between system atoms) and $\delta$ (between a system atom and environment atom) are perpendicular.
The environment atoms are treated as 'three-level' atoms: they have a ground state $\ket{g}$, a short-lived excited state $\ket{e}$ and a Rydberg state $\ket{r}\neq\ket{p},\ket{s}$.  
The excited state is coupled to the zero-temperature photonic continuum, inducing spontaneous emission with decay rate $\Gamma_p$ to the ground state $\ket{g}$.  
Both the $\ket{g} \leftrightarrow \ket{e}$ and $\ket{e} \leftrightarrow \ket{r}$ transition are optically driven, with Rabi frequencies $\Omega_p$ and $\Omega_c$ and detunings $\Delta_p$ and $\Delta_c$ respectively.
The Hamiltonian for the environment atoms, in the rotating wave approximation, is given by:
\begin{eqnarray}
\mathcal{H}_\mathrm{env} =& \sum_\alpha \left[\frac{\Omega_p}{2}\bdyad{e}{g}{\alpha} + \frac{\Omega_c}{2}\bdyad{r}{e}{\alpha} + \mathrm{H.c.}\right] - \Delta_p\bdyad{e}{e}{\alpha}  \nonumber \\ &- (\Delta_p+\Delta_c)\bdyad{r}{r}{\alpha} +\sum_{\alpha<\beta} V_{\alpha\beta}^{(rr)} \bdyad{r}{r}{\alpha}\bdyad{r}{r}{\beta}. \label{eq:Henv}
\end{eqnarray} 
The final term in the environment Hamiltonian corresponds to the inter-atomic van der Waals interaction $V_{\alpha\beta}^{(rr)}$ between atoms in Rydberg states $\ket{r}$, where $\alpha$ and $\beta$ label the environment atoms. 

The system and environment atoms interact via the Rydberg states of the environment atoms, $\ket{r}$.
These interactions depend on the state of the system atoms:
\begin{equation}
 \mathcal{H}_\mathrm{int} = \sum_{n,\alpha}{\bar{V}_{n\alpha} \dyad{\pi_n}{\pi_n}} \bdyad{r}{r}{\alpha},
\end{equation}
where $\bar{V}_{n\alpha} = V^{(pr)}_{n\alpha} + \sum_{m\neq n} V^{(sr)}_{m\alpha}$ is the strength of the interaction between the system in the state $\ket{\pi_n}$ and a specific environment atom $\alpha$ in the Rydberg state $\ket{r}$.
The interaction between system atom $n$ in state $\ket{p}$ ($\ket{s}$) and environment atom $\alpha$ in state $\ket{r}$ is given by $V^{(pr)}_{n\alpha}$ ($V^{(sr)}_{n\alpha}$).
These pairwise interactions depend strongly on distance, such that the strongest interactions are between adjacent environment and system atoms, $\alpha=n$.

For the given setup, we can obtain the system dynamics in the following manner.  We first define an initial state for our setup that is experimentally accessible and corresponds to a localized excitation: $\rho(0)=\left( \ket{\pi_1} \otimes \ket{g...g} \right) \left( \bra{\pi_1}\otimes \bra{g...g}  \right)$.  The state $\rho$ is then propagated in time according to the master equation:
\begin{equation} \label{eq:dyn}
 \frac{d}{dt}\rho(t) = -i [\mathcal{H}, \rho(t)] + \sum_\alpha \mathcal{D}[L_\alpha]\rho(t),
\end{equation}
where $\mathcal{H}= \mathcal{H}_\mathrm{sys} + \mathcal{H}_\mathrm{env} + \mathcal{H}_\mathrm{int}$ and $\mathcal{D}[L_\alpha]\rho = L_\alpha \rho L_\alpha^\dagger - \frac{1}{2}\left(L_\alpha^\dagger L_\alpha \rho + \rho L_\alpha^\dagger L_\alpha  \right)$ terms describe the effect of spontaneous emission on the setup.  Here $L_\alpha = \sqrt{\Gamma_p}\bdyad{g}{e}{\alpha}$.

We specify the inter-atomic distances, atoms and states in the setup as in~\cite{Schon18NJP013011}.
The aggregate spacing is $d=5$~$\mu$m and the aggregate-environment atom spacing is $\delta=2$~$\mu$m. 
We choose the states of the aggregate atoms to be $\ket{p}=\ket{43p}$ and $\ket{s}=\ket{43s}$ of $^{87}$Rb.
A dimer (two system atoms) then has a lifetime of approximately 56~$\mu$s~\cite{Bete09PRA052504}, which is much longer than the timescale of dynamics that we will consider.
  For the environment atoms we choose the states $\ket{g}=\ket{5s}$, $\ket{e}=\ket{5p}$ and $\ket{r}=\ket{38s}$ of $^{87}$Rb.
The decay rate from $\ket{e}$ is $\Gamma_p = 6.1$~MHz.

In this setup, we want to obtain thermalization of the system.  
This provides a resource for quantum simulation of systems in real (thermal) environments.
We are thus interested in preparing a thermal state:
\begin{equation}
 \rho^\mathrm{th}_{T_\mathrm{eff}} = \frac{1}{Z}\sum_n{e^{-E_n/(kT_\mathrm{eff})} \dyad{\varphi_n}{\varphi_n}}, 
\end{equation}
where $T_\mathrm{eff}$ is the effective temperature, $k$ is the Boltzmann constant,  and $Z = \mathrm{Tr} \{ e^{-\mathcal{H}_\mathrm{sys}/(kT_\mathrm{eff})} \}$.
The eigenstates and eigenenergies of $\mathcal{H}_\mathrm{sys}$ (equation~\ref{eq:Hfullsys}) are denoted by $\ket{\varphi_n}$ and $E_n$, respectively.

Note that the temperature scale $kT_\mathrm{eff}$ is not given by an ambient temperature of a thermal bath. The ambient temperature is typically on the order of~$\mu$K, and we also have an additional atomic component of our environment.  Here the temperature scale is given in terms of the interaction strength $W$, which determines the eigenenergies $E_n$.

We set a time $t_f$ for which thermalization should have happened to a given thermal state.  
As addressed in the introduction, control over the thermalization timescale and the temperature of the target thermal state are important aspects for simulating general open systems.  
In this work, we focus on cost landscape prediction and analysing the results.
We thus fix the thermalization timescale to be faster than a given $t_f$, where the target thermal state has a given temperature $kT_\mathrm{eff}$.
However, our approach can be extended by varying the values $t_f$ and $kT_\mathrm{eff}$.
We will comment on the choices of $t_f$ and $kT_\mathrm{eff}$ in section~\ref{sec:scan}.

In Ref.~\cite{Schon18NJP013011}, it was shown that for the given setup, thermalization to a tunable temperature can be achieved for particular choices of the laser driving parameters ($\Omega_p$, $\Omega_c$, $\Delta_p$ and $\Delta_c$).
Here, we are similarly interested in investigating thermalization of the system by controlling the laser driving parameters for the background atoms.
However, unlike in Ref.~\cite{Schon18NJP013011}, in this paper we are interested in knowing the cost landscape generally (how well thermalization can be performed over the full parameter space).
This in turn equips us to answer other physical questions about the setup (e.g. robustness of thermalization to parameter variations or scaling the system size), as described in the introduction.

\subsection{Quantifying the outcome} \label{sec:cost}

We want to quantify how well we prepare the target thermal state.  To do this, we apply the trace distance $D(\rho_S(t_f),\rho_{T_\mathrm{eff}}^{\mathrm{th}})$ to measure the distinguishability of a given state of the system $\rho_S(t_f)$ from the target thermal state $\rho_{T_\mathrm{eff}}^{\mathrm{th}}$: 
\begin{equation}
  D(\rho_S(t_f),\rho_{T_\mathrm{eff}}^{\mathrm{th}}) = \frac{1}{2}\mathrm{Tr}\left\{|\rho_S(t_f)-\rho_{T_\mathrm{eff}}^{\mathrm{th}}|\right\},
\end{equation}
with $|\rho|=\sqrt{\rho^\dagger \rho}$. 
The state $\rho_S(t_f)$ is obtained by time-propagation of our setup $\rho(t)$ (equation~(\ref{eq:dyn})), followed by tracing out the environment atoms. 
We will use $D(\rho_S(t_f),\rho_{T_\mathrm{eff}}^{\mathrm{th}})$ (denoted by $D$ for convenience) as our cost function.

Note that the cost function is based on a state comparison at a single point in time, $t_f$.
We assume that the cost quantifies how well states have thermalized to the target state.
However, the cost does not distinguish whether the propagated state is still changing in time; such a state could dynamically pass 'close' to the target state without being effectively thermalized.
In section~3 of the SI, we show that states that are not yet steady have a minimal effect in our case, validating our association of the cost with effective thermalization.
We also provide an alternate cost function in the SI that could be used in cases where this transience issue arises.

\subsection{Gaussian processes} \label{GPintro}

In this manuscript, we numerically investigate the cost landscape.
The challenge is that simulations of our setup are computationally costly.
Also, an analytic model for the cost landscape has not been found for our setup.
The ideal case - knowing the full cost landscape - is thus impractical even for a Rydberg dimer system (just two system atoms).

Using machine learning, we can gain insight into the full cost landscape through Gaussian process (GP) regression and prediction.
GP regression fits known points in the landscape (parameters with known associated costs) and can then predict the full cost landscape.
This prediction of the cost landscape attributes a Gaussian predictive distribution for the cost at any given 'test' parameter set.
From this distribution, the predicted mean cost and standard deviation can be extracted.
A functional form is assumed for the covariance of predicted costs over parameter space.  
The covariance can depend on length-scales for characteristic variation in the cost as a function of each parameter, and these length-scales can also be estimated to provide the best fit of the cost landscape by the GP regression.

The predicted cost landscape can be used within a numerical routine to balance optimization (landscape exploitation) with landscape exploration.
We are most interested in sets of parameters providing a low cost (close to 0).
However, we are not just interested in the minimal cost and its associated parameters.
We want to know every low-cost parameter region, along with its extent in parameter space.
This information requires an exploration of the cost landscape, with higher 'priority' of exploration given to regions that may have low cost values.
As explained in the SI (section~4), the predicted cost landscape can be used to identify likely low-cost regions of the cost landscape.  Similarly, the standard deviation for the predicted cost landscape can identify regions of parameter space that should be explored due to a lack of knowledge (uncertainty in the predicted costs, such that the costs could be low).
In this way, we obtain a numerical routine that uses previous \emph{runs} (simulations and GP regression) to guide the parameters for subsequent simulations.
We use the optimization package MLOOP~\cite{Wigl16SR25890,MLOOPweb}, based on the Gaussian process regression algorithm~(2.1) from~\cite{RW} and implemented in scikitlearn~\cite{Pedr11JMLR2825}.

\subsubsection{Performance measures} \label{sec:GPperfmeas}

We would like quantitative measures of how well the GP-based numerical routine performs.  
We will consider measures that compare GP-predicted regions of the cost landscape with exact calculations of these same cost-landscape regions by scanning parameter space (extrinsic measures).
These scans are only possible for limited regions of the cost landscape, and for small system sizes.
Thus, we will also consider measures that only depend on the (region of the) cost landscape predicted by the numerical routine (intrinsic measures).
In both cases, we evaluate 2D cross-sections of the cost landscape.

\paragraph{Extrinsic measures} 
The extrinsic measures allow us to analyse the accuracy of the predicted landscape: how well it matches the exact landscape.
We introduce three measures to quantify this: (a) \emph{accuracy}, (b) \emph{accuracy for $D<C_l$}, and (c) \emph{penalty}.
\begin{itemize}
\item[(a)] The \emph{accuracy} measure is the absolute difference between the scanned cross-sections and the predicted cross-sections.  The absolute difference is taken point-wise, then averaged over every point in the cross-sections. Note that the costs obtained by both parameter scans and predictions are discrete samples from the cost landscape.
\item[(b)] The \emph{accuracy for $D<C_l$} follows the same procedure as (a), but only points where the scanned cross-sections have trace distance $D$ below some low-cost threshold $C_l$ are included in the average.  This measure thus quantifies accuracy for the low-cost regions of the cost landscape, with the choice of $C_l$ discussed in section~\ref{sec:GPest}.
\item[(c)] The \emph{penalty} is a measure for how closely the real cost landscape fits within the standard deviation for the predicted cost landscape. The contribution to the \emph{penalty} for a given cost-landscape point is the absolute distance between the real cost and the standard-deviation interval about the predicted cost (this distance is 0 when the real point is within the interval). To calculate the \emph{penalty}, we take the sum of these point-wise contributions over every point in each cross-section, then average over the cross-sections. 
\end{itemize}

It is important to note that we would like to apply the routine in a regime where such extrinsic measures are not possible.  The main benefit of using Gaussian processes to predict the landscape is that this prediction can occur where scans are not feasible: in such cases, we will need to rely on intrinsic measures.

\paragraph{Intrinsic measures}
We introduce four intrinsic measures: (d) \emph{precision}, (e) \emph{precision for $D<C_l$}, (f) \emph{absolute convergence} and (g) \emph{best cost}.
\begin{itemize}
\item[(d)] The \emph{precision} is the standard deviation averaged over every point in the cross-sections.
\item[(e)] The \emph{precision for $D<C_l$} follows the same procedure as (d), but only points where the predicted cost cross-sections have trace distance $D<C_l$ are included in the average.
\item[(f)] The \emph{absolute convergence} is the absolute difference between the current and previous predicted cost landscapes (averaged over points). This measure is a function of the number of \emph{runs} performed by the numerical routine.  A single \emph{run} of the numerical routine involves selection of parameters to test, and one simulation (see section~4 of the SI for details).
We calculate the predicted cost landscape (cross-sections) to evaluate the performance measures every 20 runs, for numerical convenience.
The 'previous' predicted cost landscape is thus calculated using 20 fewer runs than the 'current' predicted cost landscape.
\item[(g)] The \emph{best cost} is the lowest cost that has been obtained: it is the optimal cost from the set of known (evaluated) points of the cost landscape.
\end{itemize}

\section{Scanning parameter space for a dimer} \label{sec:scan}

The naive approach to obtain the cost landscape is to simply scan the variable parameters.  Recall that we have four such parameters: $(\Omega_p, \Omega_c, \Delta_p, \Delta_c)$, which means an enormous number of calculations.  

As a reference for the Gaussian process approach of the following sections we perform such a scan for the simplest nontrivial setup where a scan is tractable: a Rydberg dimer.

We choose the target thermal state temperature to be given by $kT_\mathrm{eff} = 1.2~W$.  This temperature has no special significance for the thermal state, except that it is not a limiting case of nearly zero or infinite temperature (in these cases all population is in the lowest eigenstate, and the eigenstates are equally populated, respectively).  The methods we present can be applied for any temperature.

In addition, we require that this thermal state is reached before a certain time, which we choose here to be  $t_f=2$~$\mu$s. Again, as for the choice of the temperature this time is chosen arbitrarily (but is much smaller than the lifetime of  the Rydberg states of the system ($\sim 56$~$\mu$s \cite{Bete09PRA052504}).
As discussed in section~\ref{sec:systun}, by choosing a time $t_f$ we set the maximal timescale of thermalization.
We propagate the state for 2~$\mu$s and compare the resulting state with the reference thermal state as described in section~\ref{sec:cost}.

\begin{figure}[tb]
	\centering
	\includegraphics[width=8cm]{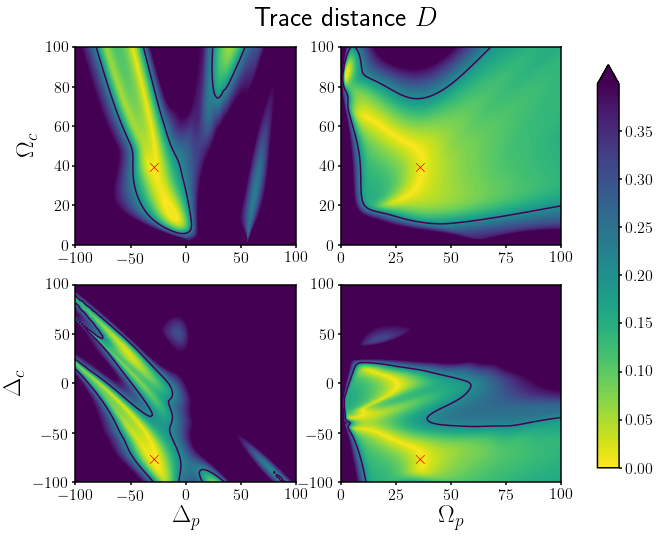}
	\caption{Parameter scan for a Rydberg dimer. 
		The cost $D$ is evaluated after 2~$\mu$s.
		The cross marks the center of the 2D cross-sections: $(\Omega_p, \Omega_c, \Delta_p, \Delta_c) = (37.1, 40.1, -26.6, -74.7)$, which has $D = 6 \times 10^{-5}$.
		The solid black lines mark the contour defined by $D=0.2$, for comparison with the cost landscapes to follow.
	}
	\label{fig:pp_scanland}
\end{figure}

\begin{figure}[tb]
	\centering
	\includegraphics[width=8cm]{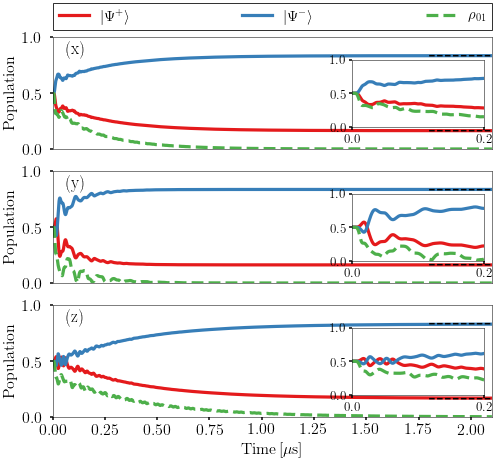}
	\caption{Population dynamics for different parameters from the 2D cost landscape cross-sections centered at $x$ (as in figure~\ref{fig:pp_scanland}). 
		 $(x)$ corresponds to the center of the cross-sections, and $(x)$, $(y)$ and $(z)$ are marked in the six cross-sections from varying each pair of parameters shown in figure~1 in the SI.
		Each point satisfies $D<0.01$.
		The eigenstates and the off-diagonal element $\rho_{01}$ are plotted over time, and black dashed lines show the target eigenstate populations.
		The target value of $\rho_{01}$ is zero.
		The inset is zoomed in on the initial dynamics of the corresponding subplot, for clarity.
	}
	\label{fig:pp_scandyn}
\end{figure}

Even though we look at just two system atoms, a full scan of the cost landscape is not feasible.
Therefore in figure~\ref{fig:pp_scanland} we present exemplary 2D cross-sections of the 4D parameter space.  We choose a low-cost central point for these cross-sections: $(\Omega_p, \Omega_c, \Delta_p, \Delta_c) = (37.1, 40.1, -26.6, -74.7)\equiv x$, which has cost $D=6 \times 10^{-5}$.  The ranges of the laser parameters are chosen to be experimental achievable~\cite{Helm16JPB,Scha14Thesis}, with detunings small enough to avoid unwanted resonances.

Each of the 2D cross-sections in figure~\ref{fig:pp_scanland} are composed of 100$\times$100 points (i.e. we evaluated 10,000 sets of parameters by time-propagating the state).  This number of points allows us to resolve features in parameter space up to one or two MHz. 
The computational time for each point in a given 2D cross-section was $\sim$~1~s (on a single core).
Many such 2D cross-sections are required to obtain information about the full 4D parameter space.

In the introduction, we raised various questions about simulation of open quantum system dynamics with Rydberg atoms, and about thermalization in particular.
For the dimer system, the cross-sections in figure~\ref{fig:pp_scanland} begin to answer some of these questions.
We can observe a range of parameters that give rise to low-cost thermalization (though still within a small subspace of the full parameter space).  We can observe the robustness of these parameters in the planes of the 2D cross-sections.  From the low-cost parameter regions, we can also sample particular sets of parameters to observe the thermalization dynamics.  As shown in figure~\ref{fig:pp_scandyn}, we found that different thermalization dynamics can be obtained from different sets of parameters.
However, note that due to the computational cost, this (scanning) approach is not scalable to larger systems, and nor can it be extended generally to explore the larger 4D parameter space even for the Rydberg dimer.

\section{Predicting the cost landscape using Gaussian processes for a dimer} \label{sec:dim}

We have seen that the computational cost is a fundamental challenge for numerical investigation of thermalization in our setup (and also more generally for simulation of scalable quantum systems).
We thus want to obtain as much information as possible from as few simulations as possible.
The information that we desire is a balance of optimization and landscape exploration: we are interested in the (multiple) regions of parameter space that give rise to low-cost thermal states, including the breadth of these regions.

\subsection{Gaussian process prediction} \label{sec:GPest}

\begin{figure*}[tb]
	\centering
	\includegraphics[width=7cm]{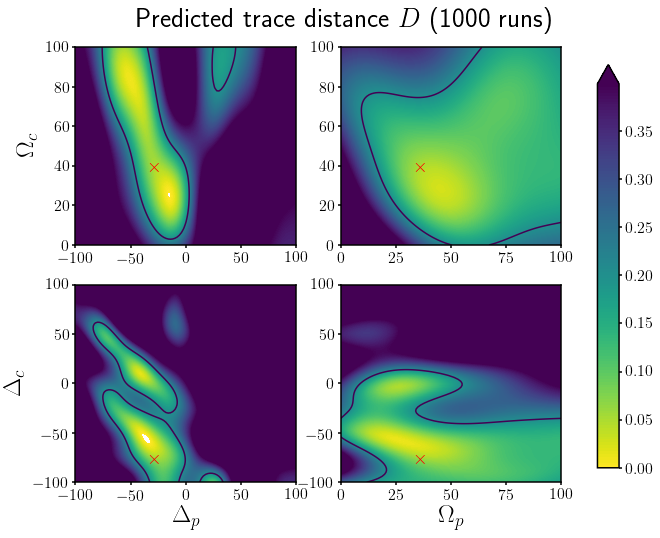}
	\includegraphics[width=7cm]{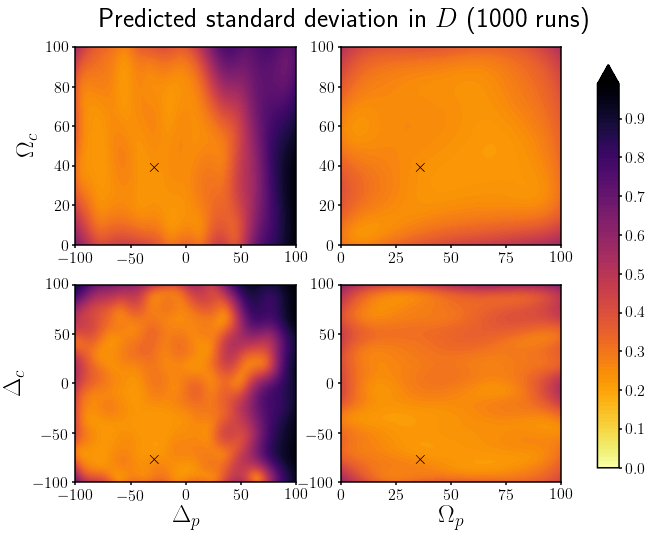}
	\caption{Predicted $D$ landscape (left) and its standard deviation (right) for thermal state preparation of a Rydberg dimer.
	The cross marks the center of the cross-sections, which is the same is in Figure~\ref{fig:pp_scanland}.
		The solid black lines mark the contour defined by $D=0.2$.
		The landscape is predicted for a 2~$\mu$s evolution time, using 1000 runs of the Gaussian process routine. 
		All four parameters were allowed to vary: for $(\Omega_p, \Omega_c, \Delta_p, \Delta_c)/(2 \pi)$~(MHz): the minimum boundary, maximum boundary and optimal parameters were $(0.1,0.1,-100.,-100.)$, $(100.,100.,100.,100.)$ and $(18.9, 92.4, 9.3, 42.5)$ respectively.  
			}
	\label{fig:pp_predland}
\end{figure*}

We use GPs, outlined in section~\ref{GPintro}, to predict the cost landscape for the Rydberg dimer setup.  We apply a numerical routine that uses a GP to both explore the cost landscape, and to exploit the landscape by focusing on low-cost regions as explained in section~4 in the SI.

In figure~\ref{fig:pp_predland} we show the predicted landscape and its associated standard deviation after 1000 runs of our numerical routine (i.e. 1000 sets of parameters were simulated).  

To compare directly with the numerical scans of the parameter space, we display the 2D cross-sections centered at the same point $x$ as the scans in figure~\ref{fig:pp_scanland}.
This comparison shows that qualitatively, the landscape prediction is very good, i.e. it identifies almost all low-cost regions.  This is remarkable since just 1000 simulations have been performed.   The four laser control parameters are allowed to vary freely within the ranges shown in the cross-sections; the parameters can be sampled from the full 4D parameter space and are not restricted to lie within the displayed cross-sections.  This is in contrast to the 100$\times$100 points for \emph{each} cross-section displayed for the parameter scans, where each of the 10,000 points lies within the displayed cross-sections.

We have noted that the parameters can vary within the full 4D space; the 2D cross-sections displayed in figure~\ref{fig:pp_predland} are not 'preferred' in any sense by the numerical routine.  For instance, the optimal set of parameters $(\Omega_p, \Omega_c, \Delta_p, \Delta_c)/(2 \pi)$~(MHz) $=$~$(18.9, 92.4, 9.3, 42.5)$, with $D<0.001$, does not lie within these cross-sections.  Since these arbitrary 2D cross-sections of the predicted 4D landscape are very good, we expect that the full 4D space is predicted similarly well. 

In figure~\ref{fig:pp_predland}, we show the predicted landscape after 1000 runs.  This number of runs is chosen based on the following considerations.
Firstly, in our implementation, prediction is more expensive for a GP with more completed runs. A possible way to circumvent this issue is described in~\cite{RW}.
Secondly, the standard deviation of the predicted cost landscape typically decreases with the number of runs, see e.g. figure~\ref{fig:pp_errormeas}.
There is thus a trade-off between the computational resources required for a GP-based numerical routine, and the precision of the predicted cost landscape.
For the dimer setup, 1000 runs is well within the 'sweet spot': firstly,  it is computationally much cheaper to use Gaussian processes than to scan parameters to investigate the cost landscape.
Secondly, the predicted landscape is very close to the actual landscape after 1000 runs: the typical \emph{accuracy}$<0.1$, as shown in figure~\ref{fig:pp_errormeas}.
Similarly, considering an intrinsic measure,  the standard deviation in the predicted cost is sufficiently small to identify likely low-cost regions of the cost landscape.
As seen in figure~\ref{fig:pp_predland}, after 1000 runs, the standard deviation for the predicted cross-sections typically takes a value between 0.2 and 0.3 for the low-cost regions.
Note that the 'sweet spot', where there is both accurate landscape prediction and a computational resource reduction from scans, grows as each simulation becomes computationally more expensive (for more details, see section~4 in the SI).

We now specify our choice of threshold $C_l$ for low-cost regions of the landscape.
We have noted that the standard deviation for the predicted cross-section is $\sim 0.2-0.3$ for regions with cost $D$ below 0.4, as seen in figure~\ref{fig:pp_predland}.
It is thus prudent to choose a value for $C_l$ close to this standard deviation value: then we include regions with predicted low cost, where the standard deviation for the prediction is around the size of the distance from zero cost (perfect thermalization). 
This way, the intrinsic likelihood (from the predicted cost) of a low cost guides our low-cost threshold.
By focusing on regions of the cost landscape with a predicted cost below 0.2, we can rule out vast regions of the cost landscape after 1000 runs.  
This is demonstrated by the contours in the predicted-cost cross-sections of figure~\ref{fig:pp_predland} (left), and we expect arbitrary cross-sections to have smaller low-cost regions (the cross-sections in figure~\ref{fig:pp_predland} are centered on a particularly low-cost point).  
We thus set the 'low-cost' threshold $C_l=0.2$.

\subsection{Quantifying prediction performance} \label{sec:dimqep}

To provide a quantitative analysis of the GP performance, we apply the extrinsic and intrinsic performance measures defined in section~\ref{sec:GPperfmeas}.  Each of these measures (aside from the \emph{best cost}) is averaged over the six 2D cross-sections produced by varying the four laser parameters (pairwise) about the central point.

\begin{figure}[tb]
	\centering
	\includegraphics[width=7cm]{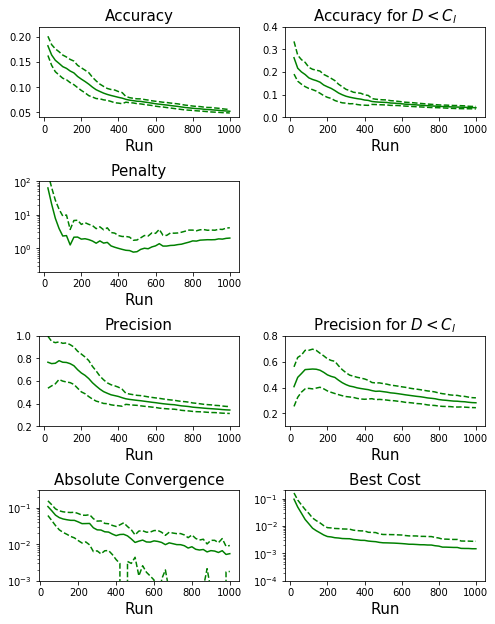}
	\caption{Performance measures for the six 2D cross-sections centered as in figures~\ref{fig:pp_scanland} and~\ref{fig:pp_predland}.
	These measures are defined in section~\ref{sec:GPperfmeas}.
	The upper three measures displayed are extrinsic, comparing the predicted cost landscape cross-sections with the scans.
	The lower four measures are intrinsic, using just the predicted cost landscape cross-sections and their standard deviations.
		The measures are calculated using 100 instances of the (stochastic) numerical routine, each of which were used to generate six 2D cross-sections of the predicted cost landscape and the associated uncertainty.
		We calculated the performance measures for each instance, and display the mean and the standard deviation of the performance measures from the resulting measure distributions.
		Note that we do not show a standard deviation below the mean in plots where this is often less than zero, for clarity.
	}
	\label{fig:pp_errormeas}
\end{figure}

Note that the numerical routine is stochastic, so the resulting landscape prediction, as in figure~\ref{fig:pp_predland}, varies for each instance of the numerical routine.  To quantify how well the GP-based numerical routine predicts the cost landscape, we have performed 100 instances of the numerical routine.  In figure~\ref{fig:pp_errormeas}, the mean over the instances is plotted for each performance measure, along with the standard deviation.  

One would expect that with more runs, i.e. more samples of the cost landscape, the landscape prediction becomes more accurate, more precisely known and approaches convergence (the landscape predictions become more similar as the number of samples increases).
In figure~\ref{fig:pp_errormeas}, this expectation is validated: the measures are almost always improving with the number of runs.  
The \emph{accuracy} (for $D<C_l$) decreases with the number of runs to a final value $\sim 0.05$ ($\sim 0.04$), which means that this is the average distance between the predicted and exact landscapes for all points (points with $D<0.2$) after 1000 runs.
The \emph{precision} and \emph{precision for $D<C_l$} measures in figure~\ref{fig:pp_errormeas} demonstrate poor initial predictions of the landscape, where the standard deviation is under-estimated prior to around 100 runs, most noticeably for the $D<C_l$ regions.  Then, after around 100 runs, the \emph{precision} measures decrease with the number of runs (the precision improves for the predicted landscape).
As expected, the \emph{absolute convergence} also demonstrates increasing consistency in the predicted landscape as the number of runs increases.

The improvement in the \emph{precision} measures, along with the improvement in \emph{accuracy}, reflects successful landscape exploration by our numerical routine.
The \emph{best cost} also improves with the number of runs: this demonstrates the optimization aspect of our numerical routine.

Recall that the \emph{penalty} grows when actual cost values lie outside the predicted standard deviation in the predicted cost.
In figure~\ref{fig:pp_errormeas}, the \emph{penalty} decreases as the prediction improves, then increases fractionally with the number of runs.  The main change in this measure is the decrease in the \emph{penalty} due to the improving landscape prediction (and more accurate standard deviation).  Initially the prediction and its standard deviation are poor due to very little information from few previous samples of the cost landscape.  The \emph{penalty} measure improves significantly within 100 runs.  The subsequent slight increase in the \emph{penalty} could be due to narrow features in the landscape that have not been sampled.  The \emph{penalty} for these features would become worse as the standard deviation in these incorrect regions of the predicted landscape is reduced with the number of runs.

If we did not have access to the scanned landscape, we would only have the intrinsic measures.  It is thus encouraging to note that the \emph{precision} measure (which averages the standard deviation in the cost landscape) is an upper bound for the \emph{accuracy} of the landscape.  This means that knowing the \emph{precision} allows us to conservatively estimate the landscape \emph{accuracy}, at least for the Rydberg dimer setup.  
The standard deviation predictions are dependent on the choice of the covariance function for the Gaussian process (as well as the optimized fitting parameters within the covariance function). 
The covariance function describes how the predicted landscape can change away from the known points in the landscape. 
We have seen that the squared-exponential covariance function~\cite{sklweb} employed by our routine (see section~4 of the SI) provides conservative standard deviation predictions.
We thus expect the same behaviour when we consider larger systems using the same covariance function.
Then the \emph{precision} could be used as a conservative estimate of the \emph{accuracy} for general system sizes.

From another intrinsic measure, the \emph{absolute convergence} measure, it appears that we could set a certain convergence threshold as a criteria to stop the numerical routine.  That is, when the predicted landscapes change less than a given amount between subsequent predictions, we might expect that the routine has converged 'close' (as determined by the threshold) to the correct cost landscape.  This approach is limited, however, when we consider individual instances of the numerical routine.  Although in figure~\ref{fig:pp_errormeas} the \emph{absolute convergence} measure (almost) monotonically decreases with the number of runs for 100 routine instances, figure~\ref{fig:pp_quaderrormeas} for the quadromer demonstrates that a single instance involves much more fluctuation in this measure.  This is due to the stochastic nature of both the sampling and the GP regression to fit the known points in the cost landscape.

The results that we have presented are for a particular choice of numerical routine, which we have found to work well for our setup.  This choice is explained in detail in section~5 of the SI, along with alternative routines.

\section{Predicting the cost landscape using Gaussian processes for a quadromer} \label{sec:quad}

We have seen that scanning parameter space does not provide an approach that can be scaled to larger system sizes within a reasonable computational time.  However, the prediction procedure using Gaussian processes performed remarkably well within a much shorter time.  We now apply this approach to gain insights from the cost landscape of a Rydberg quadromer system.

\begin{figure*}[tb]
	\centering
	\includegraphics[width=7cm]{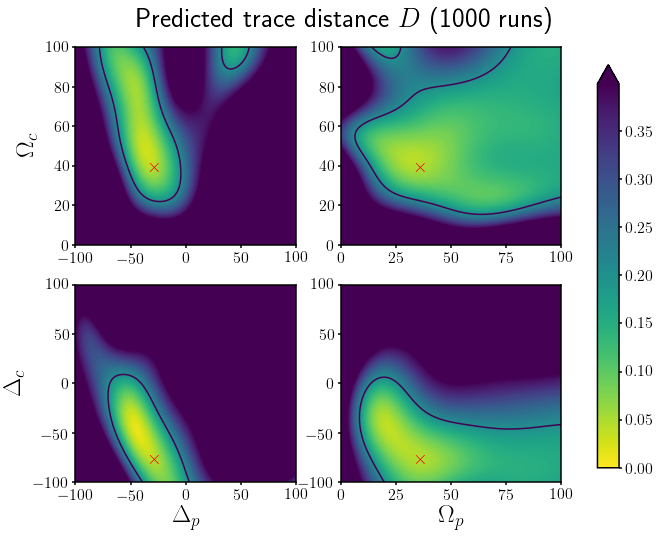}
	\includegraphics[width=7cm]{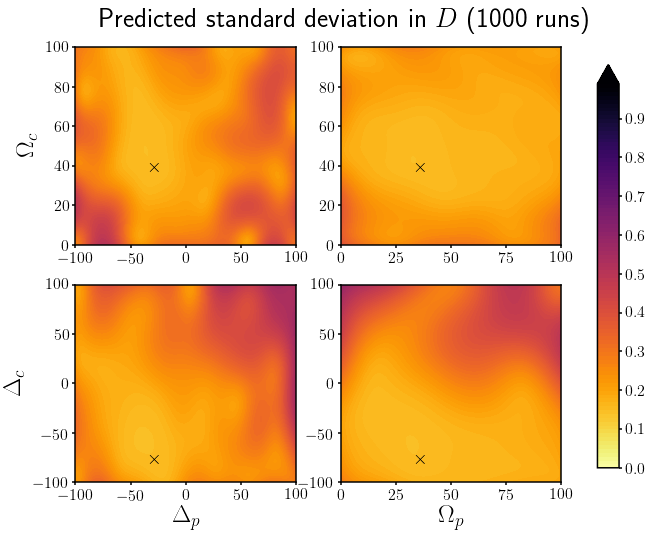}
	\caption{2D slices of the quadromer (left) predicted cost landscape and (right) its standard deviation, after 1000 runs of the numerical routine.
	The crosses mark the center of each cross-section. 
	Contours are at $D=0.2$ in the cost landscape cross-sections.
	}
	\label{fig:pp_quadpred}
\end{figure*}

\begin{figure}[tb]
	\centering
	\includegraphics[width=7cm]{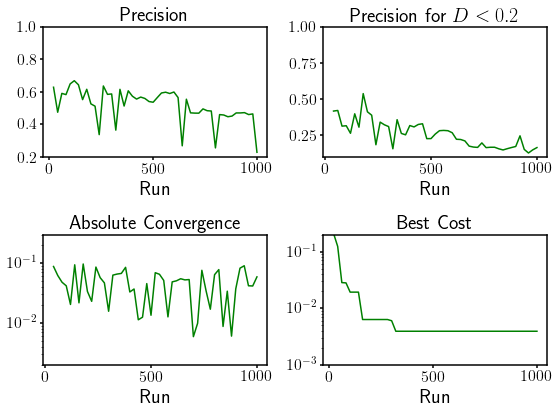}
	\caption{Intrinsic performance measures calculated using six 2D cross-sections of the cost landscape, centered at $x$.
	The six cross-sections are shown in figure~3 of the SI.}
	\label{fig:pp_quaderrormeas}
\end{figure}

We use our numerical routine to predict 2D cross-sections of the cost landscape centered at the same point $x$ as for figures~\ref{fig:pp_scanland} and~\ref{fig:pp_predland}, for comparison.
These cross-sections are presented in figure~\ref{fig:pp_quadpred}.
Here, as for the predicted cost landscape for the dimer in figure~\ref{fig:pp_predland}, 1000 points are sampled from the full 4D space to predict the landscape.

In figure~\ref{fig:pp_quadpred}, low-cost regions are predicted within the given cross-sections.  The predicted landscape gives us a sense of the robustness of these regions.
The low-cost regions are similar to those in figure~\ref{fig:pp_predland} for the dimer.
This similarity suggests that the dimer and quadromer setups possess related parameter dependence with respect to thermal state preparation.  This parameter-dependence could be extrapolated (with more data) to larger systems.  
It was found in~\cite{Schon18NJP013011} that particular laser control parameters were associated with low-cost thermal state preparation in different-sized systems. This finding was for particular points in the cost landscape of different-sized systems.  In this paper, access to the (predicted) cost landscape over large regions of parameter space allows us to explore the relationship between parameters and system size in much more detail.  For example, for the cross-sections displayed in figure~\ref{fig:pp_quadpred}, the minimal cost regions have higher cost than the dimer predictions.  We expect this increase to continue with system size, and one could investigate whether the lower cost 'peaks' for each system size can be smoothly followed through the landscape as a function of the laser parameters.

Since scanning the cost landscape is too computationally expensive, the extrinsic performance measures were not calculated for the quadromer.
This is an example of a setup where simulations are costly (a single run takes $\sim 200$~s on two cores), however a single instance of our numerical routine can be used to predict the cost landscape (as in figure~\ref{fig:pp_quadpred}).

The intrinsic performance measures were calculated using the six 2D cross-sections of the cost landscape produced by fixing four laser parameters and varying two at a time about the central point (four of the cross-sections are shown in figure~\ref{fig:pp_quadpred}, all six are shown in section~1 of the SI).  
The results are shown (as a function of the number of runs) in figure~\ref{fig:pp_quaderrormeas}.
Since the performance measures here are calculated using a single instance of the numerical routine, rather than 100 instances as in figure~\ref{fig:pp_errormeas}, we observe much more fluctuation.
Nonetheless, it can be observed from figure~\ref{fig:pp_quaderrormeas} that there is an improving trend for each measure, which is similar to the trend in figure~\ref{fig:pp_errormeas} for the dimer setup.

In figure~\ref{fig:pp_quaderrormeas}, the \emph{precision} drops to $\sim 0.23$ after 1000 runs.  
Thus, we again set the 'low-cost' threshold $C_l=0.2$.
The final value of the \emph{precision for $D<C_l$}, which is typically lower than the \emph{precision}, is $\sim 0.16$.  
As for the dimer, we expect that the \emph{precision} measures provide a conservative upper bound on the accuracy of the predicted landscape (see the discussion in section~\ref{sec:dimqep}).
Thus, we expect the predicted cost landscape for the quadromer to have \emph{accuracy} $<0.23$ overall, and we expect the \emph{accuracy for $D<C_l$} measure to be less than $0.16$.

The \emph{best cost} in figure~\ref{fig:pp_quaderrormeas} reaches a value of $\sim 4 \times 10^{-3}$ after 300 runs and remains at this value.  This is a little higher than the mean \emph{best cost} over 100 instances for the dimer case ($\sim 1.5 \times 10^{-3}$).  
Nonetheless, this \emph{best cost} value demonstrates that very-low-cost ($D \sim 0.01$) regions exist for the quadromer setup.  
These can be seen in the predicted cost landscape and standard deviation centered on the \emph{best cost}, which are shown in the SI (figure~4).

Achieving a target \emph{precision for $D<C_l$} could be used as a stopping criterion for the numerical routine (such that no more runs are performed), as it is a useful low-cost identification measure and has a relatively smooth dependence on the number of runs.
In contrast, the \emph{absolute convergence} fluctuates dramatically, which would make this measure difficult to apply as a stopping criterion.

\begin{figure*}[tb]
	\centering
	\includegraphics[width=11cm]{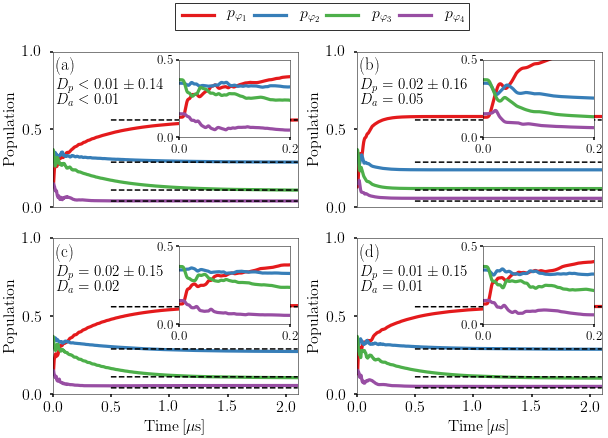}
	\caption{Quadromer dynamics plots.   
	The parameters are chosen from the predicted cost landscape cross-sections, as marked in the 2D cross-sections (figure~3 and figure~4) in the SI.
	Each set of parameters was chosen from predicted low-cost regions.
	Here the population in the $j$th lowest energy eigenstate $\ket{\varphi_j}$ is denoted by $p_{\varphi_j}$.
	We only show the eigenstate population dynamics for clarity, and the target thermal state populations are also shown (dashed lines).
	The cost compares the full state with the target thermal state; the predicted cost $D_p$ (with standard deviation) and the actual cost $D_a$ are shown for each subplot.
	The inset is a zoomed-in display of the first 0.2~$\mu$s from the respective subplot.
	}
	\label{fig:pp_quaddyn}
\end{figure*}

As we did for the dimer, we can now use the predicted cost landscape to provide a preliminary investigation of how or whether different parameters give rise to different thermalization dynamics.
In figure~\ref{fig:pp_quadpred}, as well as in figure~4 in the SI (which is centered at the \emph{best cost}), different low-cost regions are identified in the predicted cost landscape. 
We sampled points from these regions and we show their dynamics in figure~\ref{fig:pp_quaddyn}.
The four eigenstate populations undergo different dynamics, which is shown with an inset zoomed in on the initial dynamics for clarity.
The predicted costs for each point are provided with the actual costs in the figure, and the differences in each case are much lower than the standard deviation in the predicted cost (as is also the case in almost every point that we validated from the quadromer landscape). 
While the populations in the subplots $(a)$, $(c)$ and $(d)$ become steady very close to the desired eigenstate populations, the dynamics in $(b)$ are not as close (with higher cost).
Nonetheless, this point in the cost landscape (which also has a faster timescale to reach a steady-state) could be used to perform local optimization and locate a nearby lower cost with similar associated dynamics.

\section{Conclusions} \label{sec:conc}

We have tackled the general problem of extracting much information from few costly simulations.  We demonstrated that Gaussian processes perform admirably at this task in the setting of quantum simulation in atomic physics.

The successful application of GPs also provided physical insight into various questions required for simulating molecular systems.  In particular, we identified sets of parameters that give rise to thermalization, as well as the robustness of these parameter regions.  Using the parameter space information provided by GPs, we demonstrated that different thermalization dynamics can be observed in our setup.  We also obtained preliminary information about how parameters that result in thermalization vary as the system size changes.  Our method provides a useful approach for further study of this relationship.  Similarly, one could vary the temperature of the target thermal state, and observe the resulting changes in the low-cost regions of parameter space.  Our approach thus supports the development of physical insight into the controllability of our setup for molecular simulation.

It is important to note that the cost landscape could be scanned experimentally, since the experiment duration is just $2$~$\mu$s (independent of the system size), and we are interested in a parameter space described by experimentally tunable laser parameters.  
Our investigations in this paper confirmed that our setup does provide key elements required for molecular simulation, and added to the quantum simulation toolbox by identifying low-cost parameter regions, robustness, and different thermalization dynamics.  This provides a foundation and motivation for experimental exploration of the setup.

\section*{Acknowledgements}

The authors thank Jan-Michael Rost for fruitful discussions.  We acknowledge funding from the DFG: grant EI 872/4-1 through the Priority Program SPP 1929 (GiRyd).


\end{document}